\documentclass[12pt]{elsarticle}
\usepackage{graphicx}
\usepackage{amssymb}    
\usepackage{epstopdf}
\usepackage{comment}
\usepackage{xcolor}
\usepackage{hyperref}
\usepackage[final]{pdfpages}
\usepackage{frcursive}
\usepackage{bbm}
\usepackage{mathtools}
\usepackage{tabularx}
\usepackage{algorithm}
\usepackage{algorithmic}
\usepackage[normalem]{ulem}

\usepackage{upgreek}

\usepackage{yfonts}
\usepackage{lscape}
\usepackage{epsfig}
\usepackage{fullpage}
\usepackage[titletoc]{appendix}
\usepackage{fancyhdr}
\usepackage{amsmath,amssymb,xspace}
\usepackage{tabularx}
\usepackage{mathrsfs}

\newcommand{\R}{\mathbb{R}}
\newcommand{\E}{\mathbb{E}}
\newcommand{\C}{\mathbb{C}}

\newcommand{\N}{\mathbb{N}}

\newcommand{\bl}{\begin{law}}
\newcommand{\el}{\end{law}}
\newcommand{\bthm}{\begin{thm}}
\newcommand{\ethm}{\end{thm}}

\newfont{\bbb}{msbm10 scaled\magstep1}

\newtheorem{prop}{Proposition}[section]

\let \leq \leqslant
\let \geq \geqslant

\let \hat \widehat

{ \par \medskip \par
  \noindent \textit{\textbf{Demonstration\/}} : }{\null \hfill $\Box$ \par }

\makeatletter
\newcommand{\doublewidetilde}[1]{{%
  \mathpalette\double@widetilde{#1}%
}}
\newcommand{\double@widetilde}[2]{%
  \sbox\z@{$\m@th#1\widetilde{#2}$}%
  \ht\z@=.9\ht\z@
  \widetilde{\box\z@}%
}

\begin{document}

\begin{frontmatter}

  \title{From {\tt Ferminet} to PINN. Connections between neural network-based algorithms for high-dimensional Schr\"odinger Hamiltonian}
  
\author[carl]{Mashhood Khan}
\ead{alikhan4@cmail.carleton.ca}
\author[carl,crm]{Emmanuel Lorin\corref{cor1}}
\ead{elorin@math.carleton.ca}

\address[carl]{School of Mathematics and Statistics, Carleton University, Ottawa, Canada, K1S 5B6}
\address[crm]{Centre de Recherches Math\'{e}matiques, Universit\'{e} de Montr\'{e}al, Montr\'{e}al, Canada, H3T~1J4}

\cortext[cor1]{Corresponding author}

\begin{abstract}
In this note, we establish some connections between standard (data-driven) neural network-based solvers for PDE and eigenvalue problems developed on one side in the applied mathematics and engineering communities (e.g. Deep-Ritz and Physics Informed Neural Networks (PINN)), and on the other side in quantum chemistry (e.g. Variational Monte Carlo algorithms, {\tt Ferminet} (\cite{ferminet}) or {\tt Paulinet} (\cite{paulinet}) following the pioneer work of {\it Carleo et. al} \cite{carleo1,carleo2}. In particular, we re-formulate a PINN algorithm as a {\it fitting} problem with data corresponding to the solution to a standard Diffusion Monte Carlo algorithm initialized thanks to neural network-based Variational Monte Carlo. Connections at the level of the optimization algorithms are also established.
\end{abstract}

\begin{keyword} 
Scientific machine learning, neural network, Schr\"odinger Hamiltonian, optimization, variational and diffusion Monte Carlo.
\end{keyword}

\end{frontmatter}

\section{Introduction}
In less than 10 years (since 2017), Physics Informed Neural Networks (PINN) has emerged as a major player in the field of computational methods for solving direct and inverse problems involving (high-dimensional) parameterized partial differential equations (PDEs) \cite{pinns,pinns3,lagaris,PDENN}. The principle is to seek solutions to PDEs in the form of data-driven neural networks, optimized by minimizing PDE residuals. These techniques, along with others such as polynomial regression, Gaussian process, reduced order modeling, or Deep Ritz \cite{ritz}, have become highly popular in several engineering fields (e.g. mechanical engineering, fluid dynamics) as they allow to a certain extent, the tackling of problems that are inaccessible with traditional PDE solvers.

There is a tremendous effort within the mathematics and computer science communities  to address several open questions related to convergence, approximability, efficient high-dimensional optimizers, etc., in Scientific Machine Learning, including PINN-like techniques.

Independently of the development of PINN-like methods by the mathematics, engineering, computer science, and statistics communities, significant practical progress has been achieved in quantum chemistry and quantum physics in the development of neural network-based Variational Monte Carlo (VMC) and Diffusion Monte Carlo (DMC) methods. This has enabled the calculation of the electronic structure of molecules larger than those usually considered with traditional computational methods. This approach corresponds mathematically to the approximation of the lowest eigenvalues (energies) of high-dimensional Schr\"odinger operators. Following the construction of Neural Quantum States (NQS) \cite{carleo1,carleo2}, several corresponding libraries have been developed, such as {\tt Ferminet} by DeepMind$/$Google \cite{ferminet,fermi2} or {\tt Paulinet} (\cite{paulinet}). 

To our knowledge, the two above formalisms have never been unified or even rigorously connected, despite their strong similarities. In this manuscript, we establish some connections potentially relevant for the development of a mathematical formalism unifying these modern techniques. More specifically, we will reformulate the so-called Neural-Variation Diffusion Monte Carlo (N-VDMC), corresponding roughly to a standard DMC method where the trial wavefunction is computed through a VMC method (corresponding to a Deep-Ritz algorithm \cite{ritz}) as proposed for instance in {\tt Ferminet}. Let us also mention \cite{weinan}, where a neural network based variational solving for many-body Schroedinger Hamiltonian is proposed. We will also show that at the level of the optimization algorithm (gradient descent-like algorithm) a N-VDMC is indeed ``similar'' to a PINN method applied to a diffusion equation (PINN-DE) derived from the time-dependent Schr\"odinger equation. \\
In the first part of the manuscript, we will propose a short presentation of PINN, VMC, and DMC and N-VDMC algorithms. We then show that we can reformulate the PINN algorithm as a fitting problem with data, corresponding to the solution to a N-VDMC algorithm. We will then show that at the level of the optimization algorithm, a modified version of N-VDMC leads to the same gradient descent algorithm as PINN.\\
In this manuscript, we do not discuss the (very complex) structure of the neural networks usually used in quantum chemistry which follows standard strategies based on one-electron orbitals and Slater's determinants \cite{szabo}, and refer to \cite{fermi2,carleo1,carleo2} for details. However, the properties and connections presented in this manuscript are largely independent of the structure of the neural networks; the latter will be treated here as a parameterized functions.  
\subsection{Mathematical set-up}
We denote by $H$ the exact $N$-particle Schr\"odinger Hamiltonian of domain $D(H)=H^2(\R^{3N};\C)$ and by $R=(r_1,\cdots,r_N) \in \R^{3N}$ with $r_i$ the 3-dimensional coordinates of the $i$th particle, see \cite{CCT,szabo}. We denote by $\phi_G(R)$ the groundstate of $H$
\begin{eqnarray*}
  H\phi_G(R)=E_G\phi_G(R) \, ,
\end{eqnarray*}
where $H$ reads
\begin{eqnarray*}
H=-\cfrac{1}{2}\nabla^2_R + V_p(R)-E_T\, ,
  \end{eqnarray*}
and where $E_T$ is a so-called off-set energy and $V_p$ the potential energy of the considered molecule$/$atom. In this work, we consider a relatively general framework, as we will ``only'' establish {\it high-level} connections between PINN-like algorithms and neural network-based VMC and DMC algorithms.
\subsection{Basics on neural networks}
 Let us here recall the basics of fully connected neural networks. The latter are parameterized functions constructed as the composition of i) discrete convoluted linear functions, and ii) nonlinear functions. In a $d$-dimension the space, we introduce $l\in\mathbb N$, $n_i\in\mathbb N$ for $i=0, 1,\ldots, l$, with $n_0=d$. We denote by $\theta \in  \Theta:=\prod_{i=1}^l \mathbb R^{n_i\times(n_{i-1}+1)}$ the neural network parameters such that: $\theta=\{\theta_i\}_i$ with $\theta_i=(w^i,b^i) \in \R^{n_i\times n_{i-1}}\times\R^{n_i}$ and $i=1,\cdots,l$.	A network $\hat{f}$ in $\mathbb R^d$ is a function from $\prod_{i=1}^l \mathbb R^{n_i\times(n_{i-1}+1)}\times \mathbb R^d$ to $\mathbb R^{n_l}$ such that
	\begin{eqnarray}
	\hat{f}(\theta;\tau)
	&=
	w^l\sigma(w^{l-1}(\sigma(w^2\sigma(w^1\tau + b^1)+b^2)\cdots)+b^{l-1})+b^l,
	\end{eqnarray}
	where  $\sigma$ is a so-called activation function. We refer to \cite{NN1,NN2} for details on neural networks. The use of neural network-based algorithms for PDE and eigenvalue problems largely rely on approximability theorems such as  the celebrated Cybenko's Universal Approximation theorem \cite{cybenko} and more recently \cite{de2024error,de2021approximation}. Let us repeat that here, we do not discuss the very ellaborated structure of neural networks used for approximating wavefunctions in high-dimensional electronic structure problems, but refer to \cite{fermi2,carleo1,carleo2}.
\subsection{Deep-Ritz $/$ {\tt Ferminet}}
Following the celebrated works of {\it Carleo et al.} \cite{carleo1,carleo2} on neural quantum states, {\tt Ferminet} \cite{ferminet,fermi2} has become a standard electronic structure calculation library, which is based on the minimization of the Rayleigh quotient
             \begin{eqnarray*}
E_G=\min_{\phi} \cfrac{\langle H\phi,\phi\rangle}{\langle\phi,\phi\rangle} \, ,
             \end{eqnarray*}
thanks to a Variational Monte-Carlo using a neural network $\hat{\phi}(\vartheta;R)$ as ansatz; and where we have used the notation $\langle H\phi,\phi\rangle$ where $\langle\cdot ,\cdot \rangle$ denotes the $L^2$-inner product.  The groundstate $\phi_G$ is approximated by an optimized neural network $\psi_T:\, \, R\in \R^{3N} \mapsto \hat{\phi}(\vartheta^*;R)$, such that
        \begin{eqnarray*}
\vartheta^* = \textrm{argmin}_{\vartheta} \cfrac{\langle H \hat{\phi}(\vartheta;\cdot),\hat{\phi}(\vartheta;\cdot)\rangle}{\langle\hat{\phi}(\vartheta;\cdot),\hat{\phi}(\vartheta;\cdot)\rangle} \, .
        \end{eqnarray*}
        Practically, the Rayleigh quotient is minimized thanks to automatic differentiation and using Monte-Carlo integration with learning nodes $\{R^{\alpha}\}_{\alpha}$ in $\R^{3N}$ 
                \begin{eqnarray*}
\widetilde{\vartheta}^* = \textrm{argmin}_{\vartheta} \cfrac{\sum_{\alpha} H\hat{\phi}(\vartheta;R^{\alpha})\hat{\phi}(\vartheta;R^{\alpha})}{\sum_{\alpha}\hat{\phi}(\vartheta;R^{\alpha})\hat{\phi}(\vartheta;R^{\alpha})} \, ,
                \end{eqnarray*}
                hence defining an approximate groundstate.  This approach can be seen as a Deep-Ritz algorithm \cite{ritz}; corresponding to a class of {\it variational} neural network-based solvers for PDE and eigenvalue problems. In the following, the use of neural network as ansatz for minimizing the Rayleigh quotient will be referred as Neural-Variational Monte Carlo (N-VMC).
         
\subsection{Diffusion Monte Carlo and Neural Diffusion Monte Carlo}\label{subsec:dem}
Let us shortly recall the principle of the DMC method \cite{dmc0,dmc1,dmc2,dmc3,dmc4}. The latter relies on the computation over, say $Q:=\R^{3N}\times \R_+$,  of the following function
\begin{eqnarray*}
  f(R,t)=\psi_T(R)\phi(R,t) \, ,
\end{eqnarray*}
where $\psi_T$ is a given trial function approximating $\phi_G$ for a $N$-particle Schr\"odinger operator. The function $f$ satisfies the following equation on $Q$, and we will refer to the Diffusion Equation (DE) 
\begin{eqnarray}\label{EqF}
\partial_t f -\cfrac{1}{2}\nabla^2f + \nabla_R(V_0(R)f)-S_0(R)f = 0,\, \,\,\ f(R,0)=\psi^2_T(R) \, ,
\end{eqnarray}
where (for non-null $\psi_T$)
\begin{eqnarray}\label{VSE}
  \left.
  \begin{array}{rcl}
    V_0(R)& =& \nabla_R|\log\psi_T(R)| \,\,\, \textrm{(drift velocity)} \, , \\
    S_0(R)&=&E_T-E_0(R) \, \,\,\textrm{(branching term)} \, ,\\
    E_0(R)&=&\psi_T(R)^{-1}H\psi_T(R) \, \,\,\textrm{(local energy)} \, .
  \end{array}
  \right.
\end{eqnarray}
In the following, we simply rewrite \eqref{EqF} as
\begin{eqnarray}\label{EqF1}
\partial_t f + K(f) = 0,\, \,\,\ f(\cdot,0)=f_0 \, ,
\end{eqnarray}
where $f_0=\psi^2_T$ and where the operator $K$ is defined as follows.
\begin{eqnarray}\label{EqF2}
K(f) = -\cfrac{1}{2}\nabla^2f + \nabla_R(V_0(R)f)-S_0(R)f \, .
\end{eqnarray}
The groundstate and groundstate energy are obtained from
\begin{eqnarray*}
  \lim_{t\rightarrow +\infty}f(R,t)=\psi_T(R)\phi_G(R) \, ,
\end{eqnarray*}
and the so-called mixed estimator is
\begin{eqnarray*}
E_{\textrm{mix}} = \lim_{t\rightarrow +\infty} \cfrac{\langle H\phi,\psi_T\rangle}{\langle\phi,\psi_T\rangle}  = \lim_{t\rightarrow +\infty} \cfrac{\int E_0(R)f(R,t)dR}{\int f(R,t)dR} \, ,
\end{eqnarray*}
leading to $E_{G} = \langle H\phi_G,\psi_T\rangle/\langle \phi_G,\psi_T\rangle$, as long as $\psi_T$ has an overlap with $\phi_G$.  From a practical point of view the above approach is discretized using a sequence of time-dependent random walks $\{R^{\alpha}(t)\}_{\alpha}$ from $t$ to $\R^{3N}$, and allowing for the construction of a function $\widetilde{f}$ approximating $f$ in the form (see \cite{dmc4})
\begin{eqnarray*}
  \widetilde{f} \, :\, (R,t)\in Q \mapsto \cfrac{\sum_{\alpha}\widetilde{f}(R^{\alpha}(t),t)\delta(R-R^{\alpha}(t))}{\sum_{\alpha}\widetilde{f}(R^{\alpha}(t),t)} \in \R \, .
\end{eqnarray*}
A sequence of discrete times $t_0=0<t_1<\cdots<t_i<t_{i+1}<\cdots$ and $dt=t_{i+1}-t_i$ for any $i \in \N$ is introduced. For $\{R_{0}^{\alpha},t_0\}_{\alpha}$ given, we define $\widetilde{f}_i^{\alpha} = \widetilde{f}(R_{i}^{\alpha},t_{i})$, such that
\begin{eqnarray*}
  \left.
  \begin{array}{rcl}
    R_{i+1}^{\alpha} & = & R_{i}^{\alpha} +dt V_0(R_i^{\alpha}) + \sqrt{dt}\eta_i \, , \\
    \widetilde{f}_{i+1}^{\alpha} & = & \widetilde{f}_{i}^{\alpha}\exp\big(dt(S_0(R_{i+1}^{\alpha}) + S_0(R_{i}^{\alpha}))/2\big) \, ,
     \end{array}
  \right. 
\end{eqnarray*}
with $\eta_i=\eta(t_i)$ where $\eta$ is a $3N-$dimensional Gaussian random vector with null mean $\E[\eta]=0$ (in space and time) and variance equal to $dt$. Moreover $\E[\eta^2]=3N$, where $\eta^2$ denotes $\eta\cdot\eta$.  Acceptance$/$rejection process is based on Metropolis' probability \cite{dmc3} and we set for any $i$
\begin{eqnarray*}
\widetilde{f}_i(R) = \widetilde{f}(R,t_i) = \cfrac{\sum_{\alpha}\widetilde{f}_i^{\alpha}\delta(R-R_i^{\alpha})}{\sum_{\alpha}\widetilde{f}^{\alpha}_i} \, .
  \end{eqnarray*}
Finally, for $i$ large enough and denoting by $\langle \cdot \rangle_{\alpha}$ the average over the walkers, we get the mixed estimator at any time $t_i$,  $E^i_{\textrm{mix}} = \langle \hat{E}_{i-1}(R_i^{\alpha})\widetilde{f}_i^{\alpha} \rangle_{\alpha}/\langle \widetilde{f}_i^{\alpha}\rangle_{\alpha}$. 
\\
\\
\noindent{\bf Initialization}. In the N-VDMC algorithm, the trial function $\psi_T$ is computed from a neural-network VMC$/$Deep-Ritz method referred above as a N-VMC method \cite{carleo1,carleo2,ferminet,paulinet,ritz}. In this goal, we introduce a neural-network $\hat{\phi}_T(\vartheta;R)$ with parameters $\vartheta$. Deep-Ritz$/${\tt Ferminet} consists in minimizing 
\begin{eqnarray*}
\vartheta^* = \textrm{argmin}_{\vartheta}\cfrac{\langle H\hat{\phi}_T(\vartheta;R),\hat{\phi}_T(\vartheta;R)\rangle}{\langle\hat{\phi}_T(\vartheta;R),\hat{\phi}_T(\vartheta;R)\rangle} \, .
\end{eqnarray*}
We denote $f_0(R)=\hat{\phi}_T^2(\vartheta^*;R)$ which is taken as the initial data to \eqref{EqF} for both N-VDMC and PINN. In practice, for a fixed set of initial positions $\{R_0^{\alpha}\}_{\alpha}$, we define 
\begin{eqnarray*}
  \widetilde{f}_0(R) = \cfrac{\sum_{\alpha}f_0(R_0^{\alpha})\delta(R-R_{0}^{\alpha})}{\sum_{\alpha}f_0(R_0^{\alpha})} \, .
\end{eqnarray*}
  {\bf For the sake of notation simplicity, hereafter we will denote $f_i$ the DMC-solution (in place of $\widetilde{f}_i$) at time $t_i$, while $\hat{f}_i$ will refer to the PINN-solution.}
\subsection{Basics on PINN}
Physics Informed Neural Networks (PINN) \cite{pinns,pinns3,lagaris,PDENN} has become a standard methodology in engineering and applied mathematics for the data-driven computation of partial differential equations.  A PINN-like solution to a differential system ($\mathcal{S}$) is a (smooth and usually data-driven) neural network optimized thanks to the minimization of the norm of the PDE residual of ($\mathcal{S}$) evaluated at some (randomly chosen) learning nodes. This approach has recently gained a huge interest in scientific machine learning thanks to their simplicity, their nice mathematical and computational properties coming in particular from automatic differentiation and efficient stochastic gradient descent methods \cite{SGD1}. \\
        We now discuss the principle of PINN algorithms for solving \eqref{EqF}. In order to avoid a null solution (which can occur for large times), it is preferable to decompose the time-domain as follows: $\R_+=\cup_{i=1}^{\infty}[t_{i-1},t_i)$, with $t_0=0<t_1<\cdots<t_i<t_{i+1}<\cdots$ and we denote $dt=t_{i+1}-t_i$ and $Q_i=\R^{3N}\times [t_{i-1},t_{i})$. In general $dt$ does not represent a time-step (and the time-derivative is not approximated) and can often be taken much larger than usual time steps in standard computational solvers such as DMC. It is also possible to solve \eqref{EqF} by considering it as a boundary value problem, hence without approximating the time derivative. In this paper, as the objective is to propose links to N-VDMC algorithms, within PINN we propose a first order discretization of $\partial_t$, so that $dt$ actually represents the same time-step as the one used in DMC.
            \begin{itemize}
            \item On $Q_0$ and for $f_0=\psi^2_T$ given, we search for a neural network $\hat{f}:(\theta;R)\in \Theta\times \R^{3N}\rightarrow \R$ which i) optimized by minimizing the commonly called loss (objective) function
        \begin{eqnarray*}
          \mathcal{L}^{(\textrm{PINN})}_1(\theta) = \big\|\hat{f}(\theta;\cdot)-f_0 -\cfrac{dt}{2}\nabla_R^2 f_0) + dt\nabla_R(V_0f_0)-dt S_0f_0 \big\|^2_{L^2(\R^{3N})} \, ,
        \end{eqnarray*}
        where $S_0$, $V_0$ are defined in \eqref{VSE}. The approximate solution to \eqref{EqF} at time $t_1$ is denoted by $\hat{f}_1(R)=\hat{f}_1(\theta_1^*;R)$, where
          \begin{eqnarray*}
          \theta_1^* = \textrm{argmin}_{\theta}\mathcal{L}_1(\theta) \, .
          \end{eqnarray*}
        \item On $Q_i$ for $i>1$, assuming given $\hat{f}_{i-1}(R)$ (initial condition at $t=t_i$), we search for a neural network $\hat{f}:(\theta;R)\in \Theta\times \R^{3N}\rightarrow \R$ optimized by minimizing 
        \begin{eqnarray}\label{loss_i}
          \mathcal{L}^{(\textrm{PINN})}_i(\theta) = \big\|\hat{f}(\theta;\cdot) -\hat{f}_{i-1} - \cfrac{dt}{2}\nabla_R^2 \hat{f}_{i-1} + dt\nabla_R(\hat{V}_{i-1}\hat{f}_{i-1})-dt\hat{S}_{i-1}\hat{f}_{i-1} \big\|^2_{L^2(\R^{3N})} \, ,
        \end{eqnarray}
where $\hat{f}_{i-1}=\hat{\phi}^2_{i-1}$ is given and (assuming enough regularity)
\begin{eqnarray*}
  \left.
  \begin{array}{rcl}
    \hat{V}_{i-1}(R) & =& \nabla_R|\log\hat{\phi}_{i-1}(R)| \,\,\, \textrm{(drift velocity)} \, , \\
    \hat{S}_{i-1}(R)&=&E_T-\hat{E}_{i-1}(R) \, \,\,\textrm{(branching term)} \, ,\\
    \hat{E}_{i-1}(R)&=&\hat{\phi}_{i-1}^{-1}(R)H\hat{\phi}_{i-1}(R) \, \,\,\textrm{(local energy)} \, .
  \end{array}
  \right.
\end{eqnarray*}

        The approximate solution to \eqref{EqF} at $t_i$ is then denoted by $\hat{f}_i(R)=\hat{f}(\theta_i^*;R)$, where
          \begin{eqnarray*}
          \theta_i^* = \textrm{argmin}_{\theta}\mathcal{L}_i(\theta) \, .
          \end{eqnarray*}
              \end{itemize}
            In high-dimension, the $L^2-$norm are computed using Monte-Carlo integrations; this allows to connect PINN and DMC. Hence for $i$ large enough we get an approximation to $f(R,t_i)$ from which we get can obtain an approximate ground state and approximate ground state energy.
            More specifically, denoting $\{R^{\alpha}\}_{\alpha}$ a set of $M$ randomly chosen learning nodes over $\R^{3N}$ approximating \eqref{loss_i} by the commonly called {\it training loss} $\widetilde{\mathcal{L}}^{(\textrm{PINN})}_i$ defined as follows.
\begin{eqnarray}\label{approx_loss_i}
    \left.
  \begin{array}{rcl}
    \widetilde{\mathcal{L}}^{(\textrm{PINN})}_i(\theta)& = &\cfrac{1}{M}\sum_{\alpha}\big|\hat{f}(\theta;R^{\alpha}) -\hat{f}_{i-1}(R^{\alpha}) - \cfrac{dt}{2}\nabla_R^2 \hat{f}_{i-1}(R^{\alpha}) + dt\nabla_R(\hat{V}_{i-1}\hat{f}_{i-1})(R^{\alpha}) \\
    & & -dt \hat{S}_{i-1}(R^{\alpha})\hat{f}_{i-1}(R^{\alpha}) \big|^2 \, .
  \end{array}
  \right.
\end{eqnarray}
By abuse of notation, in the following we will keep the notation $\mathcal{L}^{(\textrm{PINN})}_i(\theta)$ to represent the training loss.\\
Hereafter, we will use the $L^2-$inner product ($\langle \cdot,\cdot \rangle_{L^2(\R^{3N})}$) and norm ($\|\cdot\|_{L^2(\R^{3N})}$) to define the loss functions. However when the searched solution has more regularity (say $H^s$ for $s \geq 1$, corresponding to the $L^2$-functions with derivatives (up to the $s$th ones) in $L^2$, using the $H^s$-norm can potentially improve the minimization process and$/$or provide a more precise approximate solution as recently proven in recent several papers, where PINN algorithms are applied for instance to Navier-Stokes equations, geophysical models or nonlinear Schr\"odinger equations \cite{de2024error,de2021approximation}. Further investigations will be conducted in the framework of quantum chemistry.


\subsection{Inclusion of data}
One of the key interests of neural networks is the use of data allowing to drive$/$supervise the learning of the searched eigenstate(s). It basically consists in adding to the loss function, a data-fitting contribution allowing for a more accurate neural network optimization. As an example of supervised learning, let us consider a Schr\"odinger operator with parameter-dependent potentials $V(R,p)$ with $p \in \Pi$ and $\Pi$ a bounded subset of $\R^{\pi}$ with $\pi\in \N\backslash \{0\}$.  Then we search for a neural network $\hat{\phi} \, : \, (\theta;R,p)\in \Theta\times \R^{3N}\times \Pi \mapsto \hat{\phi}(\theta;R,p) \in \R$ approximating  a $p$-dependent groundstate $\phi_G \, : \,(R,p)\in \R^{3N}\times \Pi \mapsto \phi_G (R,p) \in \R$.  Assuming that the groundstate was computed for $q$ parameters $p_{1},\cdots,p_{q}$, we add to the loss function the following contribution$/$constraint.
\begin{eqnarray*}
\mathcal{L}(\theta) \leftarrow \mathcal{L}(\theta) + \sum_{k=1}^{q}\sum_{\alpha}|\hat{\phi}(\theta;R^{\alpha},p_k)-\phi_G (R^{\alpha},p_k)|^2 \, .
  \end{eqnarray*}
As data inclusion is similarly performed within N-VDMC, Deep-Ritz$/${\tt ferminet} and PINN algorithms, in this paper we do not focus on this aspect.

\subsection{Organization}
The rest of the manuscript is organized as follows. In Section \ref{sec:comp}, we show that the standard PINN-DE algorithm can be reformulated as a fitting problem for the N-VDMC solution. Section \ref{sec:unification} is devoted to the comparison at the optimization level, of a modified N-VDMC algorithm with the standard PINN-DE algorithm. We conclude in Section \ref{sec:conclusion}.
\section{Standard N-VDMC vs PINN-DE}\label{sec:comp}
In this section, we establish connections between  N-VDMC  (trial wavefunction estimated by N-VMC) and PINN-DE algorithms. We show that a PINN-DE can be reformulated as a fitting problem for data corresponding to the DMC solution where the nodes used in the Monte Carlo integration are the DMC random walks.\\
In the following $f_i$ will refer to the DMC-solution at time $t_i$, while $\hat{f}_i$ will refer to the PINN-solution.
\subsection{Preliminary remark}
In order to connect the PINN-DE and DMC methods we will rewrite those methods in a weak form. At iteration $i\geq 1$, let us recall that the PINN loss function reads
        \begin{eqnarray*}
          \mathcal{L}^{(\textrm{PINN})}_{i}(\theta) = \big\|\hat{f}(\theta;\cdot) -\hat{f}_{i-1} - \cfrac{dt}{2}\nabla_R^2 \hat{f}_{i-1} + dt\nabla_R(\hat{V}_{i-1}\hat{f}_{i-1})-dt \hat{S}_{i-1}\hat{f}_{i-1} \big\|^2_{L^2(\R^{3N})} \, .
          \end{eqnarray*}
We denote by $\Phi$ the smooth function defined as
\begin{eqnarray*}
\Phi(\theta;\cdot)=\hat{f}(\theta;\cdot) -\hat{f}_{i-1} - \cfrac{dt}{2}\nabla_R^2 \hat{f}_{i-1} + dt\nabla_R(\hat{V}_{i-1}\hat{f}_{i-1})-dt \hat{S}_{i-1}\hat{f}_{i-1} \, ,
\end{eqnarray*}
and we rewrite $\mathcal{L}^{(\textrm{PINN})}_{i}(\theta)$ in the form
        \begin{eqnarray*}
    \left.
    \begin{array}{rcl}
      \mathcal{L}^{(\textrm{PINN})}_{i}(\theta)& =& \langle \hat{f}(\theta;\cdot) -\hat{f}_{i-1} - \cfrac{dt}{2}\nabla_R^2 \hat{f}_{i-1} + dt\nabla_R(\hat{V}_{i-1}\hat{f}_{i-1})-dt \hat{S}_{i-1}\hat{f}_{i-1},\Phi \rangle \\
&= &\langle \hat{f}(\theta;\cdot) -\hat{f}_{i-1},\Phi \rangle -dt \langle \hat{S}_{i-1}\hat{f}_{i-1}- \cfrac{dt}{2} \hat{f}_{i-1},\nabla_R^2\Phi \rangle  -dt \langle \hat{f}_{i-1}, V\nabla_R\Phi \rangle \, .
      \end{array}
\right.        
          \end{eqnarray*}
By abuse of notation (keeping the same notation for the {\it training} loss), and for randomly selecting learning nodes $\{R^{\alpha}\}_{\alpha}$, we get
        \begin{eqnarray}\label{e1}
    \left.
    \begin{array}{rcl}
      \mathcal{L}^{(\textrm{PINN})}_{i}(\theta) & = & \cfrac{1}{M}\sum_{\alpha} \big(\hat{f}(\theta;R^{\alpha}) -\hat{f}_{i-1}(R^{\alpha})\big)\Phi(R^{\alpha})  -dt \hat{S}_{i-1}(R^{\alpha})\hat{f}_{i-1}(R^{\alpha})\Phi(R^{\alpha}) \\
    & &   - \cfrac{dt}{2} \hat{f}_{i-1}(R^{\alpha})\nabla_R^2\Phi(R^{\alpha}) -dt \hat{f}_{i-1}(R^{\alpha}) \hat{V}_{i-1}(R^{\alpha})\nabla_R\Phi(R^{\alpha})\, .
      \end{array}
\right.        
          \end{eqnarray}
        Then we define $\hat{f}_{i}=\hat{f}(\theta_{i}^*;\cdot)$ where $\theta_{i}^*=\textrm{argmin}_{\theta}\mathcal{L}^{(\textrm{PINN})}_{i}(\theta)$.
        \\
\\
        Unlike PINN-DE, N-VDMC relies on the electron positions $\{R^{\alpha}_{i}\}_{\alpha}$ as follows
\begin{eqnarray*}
f_{i}(R) = \cfrac{\sum_{\alpha}f_{i}(R_{i}^{\alpha})\delta(R-R_{i}^{\alpha})}{\sum_{\alpha}f_{i}(R_{i}^{\alpha})} \, .
\end{eqnarray*}
Expanding $f_i$ about $R_{i}^{\alpha}$, we get 
\begin{eqnarray*}
    \left.
  \begin{array}{rcl}
f_i(R_i^{\alpha}) & = & f_{i}(R_{i-1}^{\alpha})\exp\big(dt(S_0(R_{i-1}^{\alpha})+S_0(R_i^{\alpha})/2)\big)\\
& = & f_{i}(R_{i-1}^{\alpha})(1 + dtS_0(R_{i-1}^{\alpha})) + dt^{3/2}\eta_{i-1}\nabla_RS_0(R^{\alpha}_{i-1}) + O(dt^2)\, .
 \end{array}
  \right.
\end{eqnarray*}
We simply denote
\begin{eqnarray}\label{fi_update}
    \left.
  \begin{array}{rcl}
f_i(R_i^{\alpha}) & = &  f_{i}(R_{i-1}^{\alpha})(1 + dtS_0(R_{i-1}^{\alpha})) + O(\eta_{i-1}dt^{3/2}) + O(dt^2)\, .
 \end{array}
  \right.
\end{eqnarray}
Moreover in the distributional sense
\begin{eqnarray*}
    \left.
  \begin{array}{rcl}
    \delta(R-R_i^{\alpha}) & = & \delta(R-R_{i-1}^{\alpha})-dt V_0(R_{i-1}^{\alpha})\delta'(R-R_{i-1}^{\alpha})+\eta_{i-1}^2\cfrac{dt}{2}\delta''(R-R_{i-1}^{\alpha})\\
    & & -\sqrt{dt}\eta_{i-1}\delta'(R-R_{i-1}^{\alpha})+ O(\eta_{i-1}dt^{3/2}) + O(dt^2) \, .
 \end{array}
  \right.
\end{eqnarray*}
Hence 
\begin{eqnarray*}
    \left.
  \begin{array}{rcl}
    f_i(R_i^{\alpha})\delta(R-R_i^{\alpha}) & = & f_{i}(R_{i-1}^{\alpha})\delta(R-R_{i-1}^{\alpha})-dt V_0(R_{i-1}^{\alpha})f_{i}(R_{i-1}^{\alpha})\delta'(R-R_{i-1}^{\alpha})\\
    & & + dtS_0(R_{i-1}^{\alpha}) f_{i}(R_{i-1}^{\alpha})\delta(R-R_{i-1}^{\alpha}) +\eta_{i-1}^2 \cfrac{dt}{2} f_{i}(R_{i-1}^{\alpha})\delta''(R-R_{i-1}^{\alpha})\\
    & & -\sqrt{dt}\eta_{i-1} f_{i}(R_{i-1}^{\alpha})\delta'(R-R_{i-1}^{\alpha}) + O(\eta_{i-1}dt^{3/2}) + O(dt^2) \, .
 \end{array}
  \right.
\end{eqnarray*}
We then deduce that for any $\Phi$ smooth
\begin{eqnarray*}
      \left.
  \begin{array}{rcl}
  \langle (f_i(R_i^{\alpha})\delta(R-R_i^{\alpha}),\Phi\rangle&  = &f_{i}(R_{i-1}^{\alpha})\Phi(R_{i-1}^{\alpha})(1+dt S_0(R_{i-1}^{\alpha})) +dtf_{i-1}(R_{i-1}^{\alpha})V_0(R_{i-1}^{\alpha})\nabla_R\Phi(R_{i-1}^{\alpha}) \\
  & & + \eta_{i-1}^2 \cfrac{dt}{2} f_{i}(R_{i-1}^{\alpha})\nabla_R^2\Phi(R_{i-1}^{\alpha}) + \eta_{i-1}\sqrt{dt}f_{i}(R_{i-1}^{\alpha})\nabla_R\Phi(R_{i-1}^{\alpha}) \\
  & & + O(\eta_{i-1}dt^{3/2}) + O(dt^2) \, .
   \end{array}
  \right.
\end{eqnarray*}
Then for any smooth function $\Phi$, we have
\begin{eqnarray}\label{e2}
        \left.
  \begin{array}{rcl}
  f_i(R_{i}^{\alpha})\Phi(R_{i}^{\alpha})& =&  f_{i}(R_{i}^{\alpha})\Phi(R_{i-1}^{\alpha})(1  + dt S_0(R_{i-1}^{\alpha}))  + dtf_{i-1}(R_{i-1}^{\alpha})V_0(R_{i-1}^{\alpha})\nabla_R\Phi(R_{i-1}^{\alpha})\\
  & & + \eta_{i-1}^2 \cfrac{dt}{2} f_{i}(R_{i-1}^{\alpha})\nabla_R^2\Phi(R_{i-1}^{\alpha}) + \eta_{i-1}\sqrt{dt}f_{i}(R_{i-1}^{\alpha})\nabla_R\Phi(R_{i-1}^{\alpha}) \\
  & & + O(\eta_{i-1}dt^{3/2}) + O(dt^2) \,.
     \end{array}
  \right.
\end{eqnarray}
In the following subsection, we will use \eqref{e2} to reformulate \eqref{e1}.
\subsection{PINN as a fitting algorithm for DMC}
The objective is here then to rewrite $\mathcal{L}^{\textrm{(PINN)}}_i$ using $\{f_i\}_i$, the N-VDMC solution. {\it At iteration $i-1$, we assume given $\hat{f}_{i-1}(R)$ the approximate solution to PINN-DE at time $t_{i}$}. We construct the smooth function $\Phi$, as follows
\begin{eqnarray*}
        \left.
        \begin{array}{lcl}
          \Phi\,:\,(\theta;R) \in \Theta\times \R^{3N} &\mapsto &\hat{f}(\theta;R) -\hat{f}_{i-1}(R)(1+ dt\hat{S}_{i-1}(R))  - \cfrac{dt}{2}\nabla_R^2 \hat{f}_{i-1}(R) \\
          & & + dt\nabla_R(\hat{V}_{i-1}(R)\hat{f}_{i-1}(R)) \, .
      \end{array}
\right.
\end{eqnarray*}
In order to connect the PINN and N-VDMC algorithms, in PINN's loss function, we select the learning nodes $\{R^{\alpha}\}_{\alpha}$ over $Q_i=[t_i,t_{i+1}]\times \R^{3N}$, as the electron positions from DMC; that is: $\{R^{\alpha}\}_{\alpha}=\{R_{i}^{\alpha}\}_{\alpha}$. For the sake of readiness, we use hereafter the following notation
\begin{eqnarray*}
        \left.
        \begin{array}{lll}
          \hat{f}_{i-1}^{\alpha} =\hat{f}_{i-1}(R_{i-1}^{\alpha}), &f_{i-1}^{\alpha} =f_{i-1}(R_{i-1}^{\alpha}),  &f_{0;i}^{\alpha}=f_0(R_i^{\alpha}),\\
          V_{i-1}^{\alpha}= \hat{V}_{i-1}(R^{\alpha}_{i-1}), & \hat{V}_{i-1}^{\alpha}\hat{f}_{i-1}^{\alpha} = (\hat{V}_{i-1}^{\alpha}\hat{f}_{i-1}^{\alpha})_{i-1}^{\alpha}, & V_{0;i-1}^{\alpha}=V_0(R^{\alpha}_{i-1})\\ 
          \nabla_R\Phi_{i}^{\alpha}(\theta)=\nabla_R\Phi(\theta;R^{\alpha}_{i}),  & \hat{S}_{i-1}^{\alpha}=\hat{S}_{i-1}(R^{\alpha}_{i-1}),   & S_{0;i-1}^{\alpha}=S_0(R^{\alpha}_{i-1}), \\
          \Phi_{i-1}^{\alpha}(\theta)=\Phi(\theta;R^{\alpha}_{i-1}) \, .& &  
               \end{array}
\right.   
  \end{eqnarray*}
We next  introduce the following sequence $\{\epsilon^{\alpha}_{i-1}\}_{\alpha}$ and $\{\delta^{\alpha}_{i-1}\}_{\alpha}$
\begin{eqnarray*}
  \epsilon^{\alpha}_{i-1}=f_{i-1}^{\alpha}  -\hat{f}_{i-1}^{\alpha},   \, \,\,  \delta^{\alpha}_{i-1}=f_{0;i-1}^{\alpha}  - \hat{f}_{i-1}^{\alpha} \, ,
\end{eqnarray*}
corresponding to the difference between i) the N-VDMC and PINN-DE solutions and ii) the trial function and the PINN-DE solution, at time $t_{i-1}$ and at $\{R_{i-1}^{\alpha}\}_{\alpha}$. Hence using $f_i^{\alpha}=f_{i-1}^{\alpha}(1+dt\hat{S}_{i-1}^{\alpha}) + O(\eta_{i-1}dt^{3/2}) + O(dt^2)$ (see \eqref{fi_update}), we get
  \begin{eqnarray}\label{DS}
        \left.
        \begin{array}{lcl}
          \hat{f}_{i-1}^{\alpha}(1+dt\hat{S}_{i-1})& = &f_{i-1}^{\alpha}(1+dtS^{\alpha}_{0;i-1})  + \epsilon_{i-1}^{\alpha}(1+dt\hat{S}^{\alpha}_{i-1}) + dtf_{i-1}^{\alpha}(\hat{S}^{\alpha}_{i-1}-S^{\alpha}_{0;i-1}) \\
          & = &f_{i}^{\alpha} - \epsilon_{i-1}^{\alpha}(1+dt\hat{S}^{\alpha}_{i-1}) + dtf_{i-1}^{\alpha}(\hat{S}^{\alpha}_{i-1}-S^{\alpha}_{0;i-1}) + O(\eta_{i-1}dt^{3/2}) + O(dt^2)\, .
          \end{array}
\right.  
  \end{eqnarray}
From \eqref{e1}, using Monte-Carlo integration $\mathcal{L}_i^{(\textrm{PINN})}$ with $\{R^{\alpha}\}_{\alpha}=\{R_{i-1}^{\alpha}\}_{\alpha}$, we then have
        \begin{eqnarray*}
        \left.
        \begin{array}{rcl}
          \mathcal{L}^{(\textrm{PINN})}_i(\theta) & = &  \cfrac{1}{M}\sum_{\alpha}\big\{\big(\hat{f}^{\alpha}(\theta;R_{i-1}^{\alpha}) -f_{i}^{\alpha}\big)\Phi^{\alpha}_{i-1}(\theta)- \cfrac{dt}{2}\hat{f}_{i-1}^{\alpha}\nabla_R^2 \Phi_{i-1}^{\alpha}(\theta) - dt \hat{V}_{i-1}^{\alpha}\hat{f}_{i-1}^{\alpha}\nabla_R\Phi_{i-1}^{\alpha}(\theta) \\
          & & + \epsilon^{\alpha}_{i-1}\Phi^{\alpha}_{i-1}(\theta)(1 + dt \hat{S}^{\alpha}_{i-1}) - dtf^{\alpha}_{i-1}(\hat{S}^{\alpha}_{i-1}-S^{\alpha}_{0;i-1})\Phi^{\alpha}_{i-1}(\theta)\big\} \, .
      \end{array}
\right.  
        \end{eqnarray*}
Then using \eqref{e2}, we can rewrite \eqref{e1} as follows 
                \begin{eqnarray}\label{tmp_loss}
        \left.
        \begin{array}{rcl}
          \mathcal{L}^{(\textrm{PINN})}_i(\theta) & = &  \cfrac{1}{M}\sum_{\alpha}\big\{\big(\hat{f}^{\alpha}(\theta;R_{i-1}^{\alpha}) -f_{i}^{\alpha}\big)\Phi^{\alpha}_{i-1}(\theta) - dt (\hat{V}_{i-1}^{\alpha}-V_{0;i-1}^{\alpha})\hat{f}_{i-1}^{\alpha}\nabla_R\Phi_{i-1}^{\alpha}(\theta) \\
          & & + \epsilon^{\alpha}_{i-1}\Phi^{\alpha}_{i-1}(\theta)\big(1 + dt \hat{S}^{\alpha}_{i-1} + \cfrac{dt}{2}\nabla^2_{i-1}\nabla_R^2 \Phi_{i-1}^{\alpha}(\theta) + dt \hat{V}_{i-1}^{\alpha}\hat{f}_{i-1}^{\alpha}\nabla_R\Phi_{i-1}^{\alpha}(\theta) \big)\\
          & & - dtf^{\alpha}_{i-1}(\hat{S}^{\alpha}_{i-1}-S^{\alpha}_{0;i-1})\Phi^{\alpha}_{i-1}(\theta)+ \eta\sqrt{dt}\hat{f}_{i-1}^{\alpha}\nabla_R\Phi_{i-1}^{\alpha}(\theta) + O(\eta_{i-1}dt^{3/2}) + O(dt^{2})\big\} \, .
      \end{array}
\right.  
                \end{eqnarray}
From \eqref{DS} and as $\nabla_R^2\hat{f}_{i-1}/2 = \nabla_R(\hat{V}_{i-1}\hat{f}_{i-1})$ (setting $\hat{f}_{i-1}=\hat{\phi}^2_{i-1}$ and {\it assuming enough regularity}) then
\begin{eqnarray*}
        \left.
        \begin{array}{rcl}
          \Phi_{i-1}^{\alpha}(\theta) & = &\hat{f}^{\alpha}(\theta;R_{i-1}^{\alpha}) -\hat{f}_{i-1}^{\alpha}(1+ dt\hat{S}_{i-1}^{\alpha}) - \cfrac{dt}{2}\nabla_R^2 \hat{f}_{i-1}^{\alpha}+ dt\nabla_R(\hat{V}_{i-1}\hat{f}_{i-1})_{i-1}^{\alpha} \\
          &= & \hat{f}^{\alpha}(\theta;R_{i-1}^{\alpha})  - f_i^{\alpha} + \epsilon_{i-1}^{\alpha}(1+dtS_{0;i-1}^{\alpha})- dtf_{i-1}^{\alpha}(\hat{S}^{\alpha}_{i-1}-S^{\alpha}_{0;i-1})+ O(\eta_{i-1}dt^{3/2}) + O(dt^{2}) \, .
      \end{array}
\right.
\end{eqnarray*}
We deduce that
\begin{eqnarray*}
        \left.
        \begin{array}{rcl}
          \big(\hat{f}^{\alpha}(\theta;R_{i-1}^{\alpha}) -f_{i}^{\alpha}\big)\Phi^{\alpha}_{i-1}(\theta) & = & \big(\hat{f}^{\alpha}(\theta;R_{i-1}^{\alpha})-f_{i}^{\alpha}\big)^2 +  \big(\hat{f}^{\alpha}(\theta;R_{i-1}^{\alpha})-f_{i}^{\alpha}\big)\\
          & & \times \big( \epsilon_{i-1}^{\alpha}(1+dtS_{0;i-1}^{\alpha})- dtf_{i-1}^{\alpha}(\hat{S}^{\alpha}_{i-1}-S^{\alpha}_{0;i-1})\big) \\
          & & + O(\eta_{i-1}dt^{3/2}) + O(dt^{2}) \, .
\end{array}
\right.
\end{eqnarray*}               
                We next use that $\eta$ is a Gaussian random process, such that
                \begin{eqnarray*}
                  \cfrac{1}{M}\sum_{\alpha} \eta_{i-1}=O\Big(\cfrac{1}{\sqrt{M}}\Big),\,\,\, \cfrac{1}{M}\sum_{\alpha} \eta^2_{i-1}=3N+ O\Big(\cfrac{1}{\sqrt{M}}\Big)\, .
\end{eqnarray*}
                This allows to simplify the expression of the loss function and getting rid of the $\eta_{i-1}$ terms, the  PINN-DE loss function can be rewritten as a fitting problem.
                     \begin{eqnarray*}
        \left.
        \begin{array}{lcl}
          \mathcal{L}^{(\textrm{PINN})}_i(\theta) & = & \cfrac{1}{M}\sum_{\alpha}\big(\hat{f}^{\alpha}(\theta;R_{i-1}^{\alpha})-f_i^{\alpha}\big)^2 + \cfrac{1}{M}\sum_{\alpha} \varepsilon_{i}^{\alpha} (\theta) + O\Big(\cfrac{1}{\sqrt{M}}\Big)\, ,
      \end{array}
\right.  
                     \end{eqnarray*}
                     where
                
                \begin{eqnarray*}
        \left.
        \begin{array}{rcl}
          \varepsilon_{i}^{\alpha} (\theta) & = & \big(\hat{f}^{\alpha}(\theta;R_{i-1}^{\alpha})-f_{i}^{\alpha}\big)\big( \epsilon_{i-1}^{\alpha}(1+dtS_{0;i-1}^{\alpha})- dtf_{i-1}^{\alpha}(\hat{S}^{\alpha}_{i-1}-S^{\alpha}_{0;i-1}) \big)\\
          & & - dt (\hat{V}_{i-1}^{\alpha}-V_{0;i-1}^{\alpha})\hat{f}_{i-1}^{\alpha}\nabla_R\Phi_{i-1}^{\alpha}(\theta)- dtf^{\alpha}_{i-1}(\hat{S}^{\alpha}_{i-1}-S^{\alpha}_{0;i-1})\Phi^{\alpha}_{i-1}(\theta)  \\
          & & + \epsilon^{\alpha}_{i-1}\Phi^{\alpha}_{i-1}(\theta)\big(1 + dt \hat{S}^{\alpha}_{i-1} + \cfrac{dt}{2}\nabla_R^2 \Phi_{i-1}^{\alpha}(\theta) + dt \hat{V}_{i-1}^{\alpha}\hat{f}_{i-1}^{\alpha}\nabla_R\Phi_{i-1}^{\alpha}(\theta) \big) + O(dt^{2}) \, .
      \end{array}
\right.  
                \end{eqnarray*}
We easily get the existence of $C_{i-1}>0$, such that
                    \begin{eqnarray*}
        \left.
        \begin{array}{rcl}
          |\varepsilon_{i}^{\alpha} (\theta)| & \leq  & \big(|\epsilon^{\alpha}_{i-1}| + C_{i-1} dt |\delta^{\alpha}_{i-1}|\big)|\hat{f}^{\alpha}(\theta;R_{i-1}^{\alpha})-f_{i}^{\alpha}| + O(dt^2) \, .
      \end{array}
\right.  
                    \end{eqnarray*}
                    We can actually refine this result.  For $i=1$, as the initial data in PINN and DMC are taken identical then $\epsilon^{\alpha}_0$ and $\delta_0^{\alpha}$ are null, so that $\hat{S}^{\alpha}_{0}=S^{\alpha}_{0;0}$, $\hat{V}^{\alpha}_{0}=V^{\alpha}_{0;0}$ and as a consequence $\varepsilon_{0}^{\alpha} (\theta)=O(dt^2)$. Hence, assuming that the loss can be as small as wanted (which is however in general, a very strong assumption), we can deduce by induction that  $\varepsilon_{i}^{\alpha} (\theta)=O(dt^2)$. Thus, PINN can be seen as a data-fitting algorithm N-VDMC from data $\{f^{\alpha}_i\}$ up to a (small) correction $\varepsilon_{i-1}^{\alpha}$. More specifically, we can state the following proposition.
                    
           \begin{prop}
             Let us denote by $\{f_i^{\alpha}\}_{\alpha}$ the N-VDMC solution at time iteration $i \geq 1$. Assume that the PINN training loss function can be as small as possible up to iteration $i-1$. Then, at iteration $i$, the PINN training loss can be reformulated as a fitting problem on $\{f_i^{\alpha}\}_{\alpha}$:
                    \begin{eqnarray*}
        \left.
        \begin{array}{lcl}
          \mathcal{L}_i(\theta) & = & \cfrac{1}{M}\sum_{\alpha}\big(\hat{f}^{\alpha}(\theta;R_{i-1}^{\alpha})-f_i^{\alpha}\big)^2 + \cfrac{1}{M}\sum_{\alpha} \varepsilon_{i}^{\alpha} (\theta) + O\Big(\cfrac{1}{\sqrt{M}}\Big)\, ,
      \end{array}
\right.  
                     \end{eqnarray*}
             where $\varepsilon_{i}^{\alpha} (\theta)=O(dt^2)$ and $M$ is the number of learning nodes.
              \end{prop}

            \section{Optimization algorithm for Modified N-VDMC and PINN-DE}\label{sec:unification}
            In this section we show that both the N-VDMC and PINN-DE algorithms lead at first order, to comparable optimization algorithms; hence lead to comparable optimized neural networks approximating the groundstate to the Schr\"odinger operator. More specifically, we show that the PINN method i) applied to the Imaginary Time Method (ITM) referred here as PINN-ITM, and ii) applied to the diffusion equation \eqref{EqF} PINN-DE, lead to a gradient descent algorithm similar to the one of a {\it modified} N-VDMC method. In the previous section, the N-VDMC algorithm \eqref{subsec:dem} is initialized once for all at $t=t_0$ thanks to a N-VMC algorithm ({\tt Ferminet}$/$Deep-Ritz). Hereafter, {\it on each} time interval $[t_i,t_{i+1}]$, the so-called modified N-VDMC is reinitialized (at $t=t_i$) from a N-VMC algorithm, {\it as long as} $\hat{\phi}_{i-1}$ is smooth and strictly positive; we refer to Section \ref{sec:comp}. Moreover within the PINN algorithms, the learning nodes are taken as the electron positions from DMC. For simplicity, in the following we also set $E_T=0$. \\
            On $Q_i$ and setting $\hat{f}(\theta;t_i)=\hat{f}_{i-1}=\hat{\phi}^2_{i-1}$ and $\hat{f}(\theta;\cdot)=\hat{\psi}(\theta;\cdot)\hat{\phi}_{i-1}$, we rewrite
            \begin{eqnarray*}
              \hat{f}(\theta;\cdot) = \hat{f}_{i-1}-dtK(\hat{f}(\theta;\cdot)) \, ,
              \end{eqnarray*}
as follows
         \begin{eqnarray*}
\left.
\begin{array}{lcl}
\hat{\psi}\hat{\phi}_{i-1} & = & \hat{\phi}^2_{i-1} + \cfrac{dt}{2}\nabla^2_R(\hat{\psi}\hat{\phi}_{i-1}) - dt\nabla_R\big(\hat{\psi}\hat{\phi}_{i-1}\nabla_R|\log\hat{\phi}_{i-1}|\big) - dt\hat{\psi}H \hat{\phi}_{i-1} \, .
\end{array}
\right.
         \end{eqnarray*}
Then we get
         \begin{eqnarray*}
\left.
\begin{array}{lcl}
  \hat{\psi}\hat{\phi}_{i-1} & = & \hat{\phi}^2_{i-1} + \hat{\psi}\cfrac{dt}{2}\nabla^2_R\hat{\phi}_{i-1} + \cfrac{dt}{2}\hat{\phi}_{i-1}\nabla^2_R\hat{\psi} + dt\nabla_R\hat{\psi}\nabla_R\hat{\phi}_{i-1}-dt\nabla_R(\hat{\psi}\nabla_R\hat{\phi}_{i-1})  - dt\hat{\psi}H \hat{\phi}_{i-1}\, ,
\end{array}
\right.
         \end{eqnarray*}
         then
         \begin{eqnarray}\label{dmc_new}
\left.
\begin{array}{lcl}
  \hat{\psi}\hat{\phi}_{i-1} & = & \hat{\phi}^2_{i-1} +\cfrac{dt}{2}\hat{\phi}_{i-1}\nabla^2_R\hat{\psi} -  \cfrac{dt}{2}\hat{\psi}\nabla^2_R\hat{\phi}_{i-1} - dt\hat{\phi}_{i-1}V_p(R)\hat{\psi} +  \cfrac{dt}{2}\hat{\psi}\nabla^2_R\hat{\phi}_{i-1}\\
  & =& \hat{\phi}_{i-1}\big(\hat{\phi}_{i-1} - dt H\hat{\psi}\big) \,.
\end{array}
\right.
         \end{eqnarray}         

            \subsection{Neural network optimization}
            In this subsection, we derive the optimization algorithm for the 3 different approaches: PINN-ITM, PINN-DE and N-VDMC. We here assume that on each $Q_i$ the same learning nodes$/$randoms walks $\{R^{\alpha}\}_{\alpha}=\{R_i^{\alpha}\}_{\alpha}$ are used in the 3 different algorithms. 
     \begin{itemize}       
\item {\bf Neural network optimization for PINN-ITM}. Let us first recall that the imaginary time method can be reformulated as a minimizing the energy associated to the Hamiltonian $H$.
            \begin{equation}\label{dennrj}
             E_G:=\min_{\chi \in L^2(\R^{3N};\R)}\mathcal{R}(\chi), \,\, \textrm{ where } \,\, \mathcal{R}(\chi)=\cfrac{\langle H\chi,\chi\rangle}{\langle \chi,\chi\rangle} \, .
            \end{equation}
            Minimizing $\mathcal{R}$ the Rayleigh quotient, is equivalent under technical conditions to the imaginary time method. From for any $t$ and assuming given a normalized function $\phi_0$ in $\R^{3N}$, we search for $\psi$ in $Q=\R^{3N}\times\R_+$ solution to the time-dependent Schr\"odinger equation in imaginary time, such that
      \begin{eqnarray}\label{DDT1}
\left.
\begin{array}{lcl}
\partial_t \psi(R,t) & =& -H\psi(R,t), \, \, \, (R,t) \in Q, \\
\psi(R,0)& = & \phi_0(R),  \, \, \, R\in \R^{3N},\\
\phi(R,t) & = & \cfrac{\psi(R,t)}{\|\psi(\cdot,t)\|_{L^2(\R^{3N};\R)}},  \, \, \,(R,t) \in Q\, .
\end{array}
\right.
\end{eqnarray}      
      By showing 
      \begin{eqnarray*}
\cfrac{d}{dt}E\Big(\cfrac{\psi(\cdot,t)}{\|\psi(\cdot,t)\|_{L^2(\R^{3N};\R)}}\Big) \leq 0 \, ,
        \end{eqnarray*}
      for negative potentials (see \cite{bao,baocai}), then $\lim_{t\rightarrow \infty}\psi(\cdot,t) = \phi_G$. In other words  under technical assumptions, solving \eqref{dennrj} is equivalent to solving \eqref{DDT1}. \\
      In practice, in order to solve \eqref{DDT1} the imaginary time method requires a time-discretization $t_0=0<t_1<\cdots<t_i<t_{i+1}<\cdots$. Then, assumed given a normalized function $\phi_{i-1}$ in $\R^{3N}$, we search for $\phi_{i}$ in $\R^{3N}$ such that on $Q_i=\R^{3N}\times [t_{i-1},t_{i}]$
       \begin{eqnarray}\label{DDT2}     
\left.
\begin{array}{lcl}
\partial_t \psi(R,t) & =& -H\psi(R,t), \, \, \, (R,t) \in Q_i, \\
\psi(R,t_{i-1})& = & \phi_{i-1}(R),  \, \, \, R\in \R^{3N},\\
\phi_{i}(R) & = & \cfrac{\psi(R,t_{i})}{\|\psi(\cdot,t_{i})\|_{L^2(\R^{3N})}},  \, \, \,R \in \R^{3N}\, .
\end{array}
\right.
\end{eqnarray}
Semi-discretizing \eqref{DDT2} in time, we get
       \begin{eqnarray}\label{DDT2bis}     
\left.
\begin{array}{lcl}
\psi_{i}(R) & =& \phi_{i-1}(R) - dt H\phi_{i-1}(R), \, \, \, R \in \R^{3N}, \\
\phi_{i}(R) & = & \cfrac{\psi_i(R)}{\|\psi_i\|_{L^2(\R^{3N})}},  \, \, \,R \in \R^{3N}\, .
\end{array}
\right.
\end{eqnarray}
       Solving \eqref{DDT2bis} with PINN is equivalent to searching for a neural network using PINN $\hat{\psi}(\overline{\vartheta}_i;R)$ such that $\overline{\vartheta}_i=\textrm{argmin}_{\vartheta}\mathcal{M}_i(\vartheta)$, where (at the continuous level)
            \begin{eqnarray*}
                 \left.
          \begin{array}{lcl}
            \mathcal{M}^{(\textrm{PINN})}_i(\vartheta)  & = & \big\|\hat{\psi}(\vartheta;\cdot) -(1-dt H)\hat{\phi}_{i-1}\big\|^2_{L^2(\R^{3N})} \, .
                 \end{array}
          \right.    
            \end{eqnarray*}
            Hence
            \begin{eqnarray*}
                               \left.
          \begin{array}{lcl}
       \nabla_{\vartheta} \mathcal{M}^{(\textrm{PINN})}_i(\vartheta) & = & 2 \big\langle \nabla_{\vartheta}\hat{\psi}(\vartheta;\cdot),\hat{\psi}(\vartheta;\cdot)-(1-dt H)\hat{\phi}_{i-1} \big\rangle \, .    
                        \end{array}
          \right.  
            \end{eqnarray*}
            Thus, we get
            \begin{prop}
           The gradient descent algorithm applied to the training loss approximating the continuous PINN-ITM loss $ \mathcal{M}^{(\textrm{PINN})}_i$,  using learning nodes $\{R_i^{\alpha}\}_{\alpha}$ on $\R^{3N}$, reads as follows
           \begin{eqnarray}
             \label{loss1}
             \vartheta_{k+1} = \vartheta_k - \cfrac{2\nu_k}{M}\sum_{\alpha}\big\{\nabla_{\vartheta}\hat{\psi}(\vartheta_k;R_{i}^{\alpha})\big[\hat{\psi}(\vartheta;R_{i}^{\alpha})-\hat{\phi}_{i-1}(R_{i}^{\alpha}) + dt H\hat{\phi}_{i-1}(R_{i}^{\alpha})\big]\big\} \, ,
           \end{eqnarray}
           with learning rate $\nu_k$.
\end{prop}

\item {\bf Neural network optimization for PINN-DE.} We here reformulate the DE presented in Subsection \ref{subsec:dem} from \eqref{DDT2} on $Q_i$ by multiplying the PDE by $\phi_{i-1}$. We define $f := \phi_{i-1}\psi$ which satisfies on $Q_i$     
\begin{eqnarray}\label{DDT3}     
\left.
\begin{array}{lcl}
\partial_t f(R,t) & =& -K(f_{i-1})(R), \, \, \, (R,t) \in Q_i, \\
f(R,t_{i-1})& = & f_{i-1}(R),  \, \, \, R\in \R^{3N} \, .
\end{array}
\right.
\end{eqnarray}      
Then assuming $\hat{f}_{i-1}$ given,  a PINN method consists in searching for a neural network $\hat{f}(\theta_i^*;\cdot)$ with $\theta^*_i=\textrm{argmin}_{\theta}\mathcal{L}^{(\textrm{PINN})}_i(\theta)$, where (at the continuous level)
     \begin{eqnarray*}
                 \left.
          \begin{array}{lcl}
            \mathcal{L}^{(\textrm{PINN})}_i(\theta)  & = & \big\|\hat{f}(\theta;\cdot) -\hat{f}_{i-1} - \cfrac{dt}{2}\nabla_R^2 \hat{f}_{i-1} + dt\nabla_R(\hat{V}_{i-1}\hat{f}_{i-1})-dt\hat{S}_{i-1}\hat{f}_{i-1}\big\|^2_{L^2(\R^{3N})} \\
            & = & \big\|\hat{f}(\theta;\cdot)  -\hat{f}_{i-1} + dt K(\hat{f}_{i-1}) \big\|^2_{L^2(\R^{3N})}
                 \end{array}
          \right.    
          \end{eqnarray*}
     The gradient descent method requires the gradient of $\mathcal{L}^{(\textrm{PINN})}_i$
     \begin{eqnarray*}
       \left.
       \begin{array}{lcl}
       \nabla_{\theta} \mathcal{L}^{(\textrm{PINN})}_i(\theta) & = & 2 \big\langle \nabla_{\theta}\hat{f}(\theta;\cdot), \hat{f}(\theta;\cdot)  -\hat{f}_{i-1} +dt K(\hat{f}_{i-1})\big\rangle \, .
                 \end{array}
       \right.
     \end{eqnarray*}
     If we set
     \begin{eqnarray*}
       \hat{f}(\theta;\cdot)=\hat{\phi}_{i-1}\hat{\phi}(\theta;\cdot), \,\,\, \hat{f}_{i-1}=\hat{\phi}_{i-1}^2 \, ,
       \end{eqnarray*}
then, from \eqref{dmc_new} we get
     \begin{eqnarray*}
       \left.
       \begin{array}{lcl}
         \nabla_{\theta} \mathcal{L}^{(\textrm{PINN})}_i(\theta) & = & 2 \big\langle \hat{\phi}_{i-1}\nabla_{\theta}\hat{\phi}(\theta;\cdot), \hat{\phi}_{i-1}\hat{\phi}(\theta;\cdot)  -\hat{\phi}^2_{i-1} + dt K(\hat{\phi}^2_{i-1})\big\rangle \\
         &  =& 2 \big\langle \hat{\phi}^2_{i-1}\nabla_{\theta}\hat{\phi}(\theta;\cdot),\hat{\phi}(\theta;\cdot)  -\hat{\phi}_{i-1} + dtH \hat{\phi}_{i-1} \big\rangle \, .
         
                 \end{array}
       \right.
     \end{eqnarray*}
Setting $\langle u,v \rangle_{\hat{\phi}_{i-1}} = \langle \hat{\phi}^2_{i-1} u,v \rangle$, we can in particular we rewrite the loss function as follows
          \begin{eqnarray*}
       \left.
       \begin{array}{lcl}
         \nabla_{\theta} \mathcal{L}^{(\textrm{PINN})}_i(\theta)  &  =& 2 \big\langle \nabla_{\theta}\hat{\phi}(\theta;\cdot),\hat{\phi}(\theta;\cdot)  -\hat{\phi}_{i-1} + dtH \hat{\phi}_{i-1} \big\rangle_{\hat{\phi}_{i-1}} \, .
                 \end{array}
       \right.
          \end{eqnarray*}
            Hence, we get
            \begin{prop}
              The gradient descent algorithm applied to the training loss approximating the continuous PINN-DE loss $ \mathcal{L}^{(\textrm{PINN})}_i$, using learning nodes $\{R_i^{\alpha}\}_{\alpha}$ on $\R^{3N}$, reads as follows
       \begin{eqnarray}\label{loss2}
     \theta_{k+1} = \theta_k - \cfrac{2\nu_k}{M}\sum_{\alpha}\big\{\sum_{\alpha}\hat{\phi}^2_{i-1}(R_i^{\alpha}) \nabla_{\theta}\hat{\phi}(\theta;R_i^{\alpha})\big[\hat{\phi}(\theta;(R_i^{\alpha}))  -\hat{\phi}_{i-1}(R_i^{\alpha}) + dtH \hat{\phi}_{i-1}(R_i^{\alpha})\big]\big\} \, ,
       \end{eqnarray}
              with learning rate $\nu_k$.    
\end{prop}
      
            \item {\bf Neural network optimization for N-VDMC}. The N-VDMC method is a combination of N-VCM and DMC methods. More specifically, unlike the ``standard'' N-VDMC presented in the previous section where N-VMC was only applied once at $t=t_0$, we here assume that VMC is applied at each $t_i$ with $i \geq 1$. In other words, on $Q_i$, we initialize the wavefunction with a N-VMC algorithm ({\tt Ferminet}), as long as the initial states $\hat{\phi}_{i-1}$ are smooth and positive. Then we apply a DMC algorithm from $t_{i-1}$ to $t_{i}$. More specifically, we minimize $\mathcal{R}_i(\hat{\psi}(\vartheta))$ where
     \begin{eqnarray*}
 \mathcal{R}_i(\hat{\psi}(\vartheta)) = \cfrac{\langle H\hat{\psi}(\vartheta;\cdot),\hat{\psi}(\vartheta;\cdot)\rangle}{\langle \hat{\psi}(\vartheta;\cdot),\hat{\psi}(\vartheta;\cdot)\rangle} \, ,
     \end{eqnarray*}
    and where the corresponding gradient descent method is initialized with the solution to the DMC solution at time $t_{i-1}$ (which differs from $\hat{\phi}_{i-1}$). The converged solution is then denoted by $\hat{\phi}_{i-1}$. Let us consider VMC$/$DMC on $Q_i$ assuming that $\hat{\psi}$ (from DMC) is such that
            \begin{eqnarray*}
                 \left.
          \begin{array}{lcl}
            \hat{\psi}(\vartheta;\cdot) & = & \hat{\phi}_{i-1}- \cfrac{dt}{2}H\hat{\psi}(\vartheta;\cdot) -\cfrac{dt}{2} H\hat{\phi}_{i-1} \, ,
                 \end{array}
          \right.    
            \end{eqnarray*}
            and $\langle \hat{\phi}_{i-1},\hat{\phi}_{i-1}\rangle=1$.

            We have (for $dt$ small enough)
     \begin{eqnarray*}
                          \left.
          \begin{array}{lcl}           
            \langle \psi(\vartheta;\cdot),\psi(\vartheta;\cdot)\rangle & = &1 - dt\langle \hat{\phi}_{i-1},H(\psi(\vartheta;\cdot)+\hat{\phi}_{i-1})\rangle + \cfrac{dt^2}{4}\|H\big(\psi(\vartheta;\cdot)+\hat{\phi}_{i-1}\big)\|^2\\
            & = & 1-dt\langle H\hat{\phi}_{i-1},\hat{\phi}_{i-1}\rangle -dt\langle \psi(\vartheta;\cdot), H\hat{\phi}_{i-1}\rangle + O(dt^2) \\
               & = & 1-2dt\langle \psi(\vartheta;\cdot), H\hat{\phi}_{i-1}\rangle + O(dt^2) \, .         
                 \end{array}
          \right.    
            \end{eqnarray*}     
     Then $\nabla_{\theta}\langle \psi(\vartheta;\cdot),\psi(\vartheta;\cdot)\rangle  = -dt \langle \nabla_{\theta}\psi(\vartheta;\cdot),H\hat{\phi}_{i-1}\rangle + O(dt^2)$.  Hence
             \begin{eqnarray*}
                 \left.
          \begin{array}{lcl}
            \cfrac{\big\langle H\hat{\psi}(\vartheta;\cdot),\hat{\psi}(\vartheta;\cdot)\big\rangle}{\big\langle \psi(\vartheta;\cdot),\psi(\vartheta;\cdot)\big\rangle} &  = &\cfrac{2}{dt}\cfrac{\big\langle \hat{\phi}_{i-1}-\hat{\psi}(\vartheta;\cdot),\hat{\psi}(\vartheta;\cdot)\big\rangle}{\big\langle \psi(\vartheta;\cdot),\psi(\vartheta;\cdot)\big\rangle}  -\cfrac{\big\langle H\hat{\phi}_{i-1} , \hat{\psi}(\vartheta;\cdot)\big\rangle}{\big\langle \psi(\vartheta;\cdot),\psi(\vartheta;\cdot)\big\rangle} \\
& = & \cfrac{2}{dt}\Big(\cfrac{\big\langle \hat{\phi}_{i-1},\psi(\vartheta;\cdot)\big\rangle}{\big\langle \psi(\vartheta;\cdot),\psi(\vartheta;\cdot)\big\rangle}-1\Big)  -\cfrac{\big\langle \hat{\psi}(\vartheta;\cdot),H\hat{\phi}_{i-1} \big\rangle}{\big\langle \psi(\vartheta;\cdot),\psi(\vartheta;\cdot)\big\rangle} \\
            & = & \cfrac{2}{dt}\Big(\cfrac{1-dt\big\langle \psi(\vartheta;\cdot),H\hat{\phi}_{i-1}\big\rangle/2 -dt\big\langle H\hat{\phi}_{i-1},\hat{\phi}_{i-1}\big\rangle/2 }{\big\langle \psi(\vartheta;\cdot),\psi(\vartheta;\cdot)\big\rangle}-1\Big) \\
            & & -\cfrac{\big\langle \hat{\psi}(\vartheta;\cdot),H\hat{\phi}_{i-1} \big\rangle}{\big\langle \psi(\vartheta;\cdot),\psi(\vartheta;\cdot)\big\rangle} \\
& = & \cfrac{2}{dt}\Big(\cfrac{1}{\big\langle \psi(\vartheta;\cdot),\psi(\vartheta;\cdot)\big\rangle}-1\Big)  -2\cfrac{\big\langle \hat{\psi}(\vartheta;\cdot),H\hat{\phi}_{i-1} \big\rangle}{\big\langle \psi(\vartheta;\cdot),\psi(\vartheta;\cdot)\big\rangle}- \cfrac{\big\langle H\hat{\phi}_{i-1},\hat{\phi}_{i-1}\big\rangle}{\big\langle \psi(\vartheta;\cdot),\psi(\vartheta;\cdot)\big\rangle} \, .
                 \end{array}
          \right.    
             \end{eqnarray*}

 Then
             \begin{eqnarray*}
                 \left.
          \begin{array}{lcl}
            \cfrac{\big\langle H\hat{\psi}(\vartheta;\cdot),\hat{\psi}(\vartheta;\cdot)\big\rangle}{\big\langle \psi(\vartheta;\cdot),\psi(\vartheta;\cdot)\big\rangle} &  = & 4\langle \psi(\vartheta;\cdot), H\hat{\phi}_{i-1}\rangle  -2\big\langle \hat{\psi}(\vartheta;\cdot),H\hat{\phi}_{i-1} \big\rangle \\
            & & -\big\langle H\hat{\phi}_{i-1},\hat{\phi}_{i-1}\big\rangle  + O(dt) \\
            & = &  2\big\langle \hat{\psi}(\vartheta;\cdot),H\hat{\phi}_{i-1} \big\rangle  - \langle H\hat{\phi}_{i-1},\hat{\phi}_{i-1}\rangle  + O(dt) \, .
                 \end{array}
          \right.    
             \end{eqnarray*}

Hence the minimization of $\mathcal{R}$ is obtained from
\begin{eqnarray*}
                   \left.
          \begin{array}{lcl}
            \nabla_{\theta}\mathcal{R}_i(\vartheta) &= & 2\big\langle\nabla_{\vartheta}\hat{\psi}(\vartheta;\cdot),H\hat{\phi}_{i-1}\big\rangle +O(dt) \\
            &= & -\cfrac{2}{dt}\big\langle\nabla_{\vartheta}\hat{\psi}(\vartheta;\cdot),\hat{\psi}(\vartheta;\cdot)-\hat{\phi}_{i-1}+dtH\hat{\psi}(\vartheta;\cdot)\big\rangle   +O(dt)\, .
\end{array}
          \right.     
     \end{eqnarray*}
Thus we can state the following proposition.
              \begin{prop}
            The gradient descent algorithm applied to N-VDMC where the continuous loss function $\mathcal{R}_i$ is approximated thanks to the learning nodes $\{R_i^{\alpha}\}_{\alpha}$ on $\R^{3N}$, reads as follows
      \begin{eqnarray}\label{loss3}
     \vartheta_{k+1}  =\vartheta_k - \cfrac{2\nu_k}{M}\sum_{\alpha}\big\{\nabla_{\vartheta}\hat{\psi}(\vartheta_k;R_{i}^{\alpha})\big[\hat{\psi}(\vartheta_k;R_{i}^{\alpha})-\hat{\phi}_{i-1}(R_{i}^{\alpha}) + dt H\hat{\phi}_{i-1}(R_{i}^{\alpha}) +O(dt^2) \big]\big\}\, ,
      \end{eqnarray}
                with learning rate $\nu_k$. 
              \end{prop}
              \end{itemize}
\subsection{Conclusion}
The above discussion allows to deduce that up to a $O(dt^2)$, the PINN-DE and N-VDMC algorithms lead to the same optimization algorithms \eqref{loss2} and \eqref{loss3}. In addition, by choosing $\langle \cdot ,\cdot \rangle_{\hat{\phi}_{i-1}}$ as the inner product (at iteration $i$), still up to a $O(dt^2)$ the PINN-ITM algorithm also leads to the same optimization algorithm, see \eqref{loss1}.\\
\\
Recall that the ``direct'' N-VDMC as described in Section \ref{subsec:dem} (see \cite{dmc4}) and unlike the modified version used above, does not require to the use of VMC at each time iteration $t_i$ and as a consequence is much less computationally complex. Notice that PINN-like algorithms do not necessarily require the discretization of the time-derivative. Instead we can search for a neural network on $Q_T=\R^{3N}\times[0,T]$ (for $T$ large enough) with $f_0=\psi^2_T$ given, for a neural network $\hat{f}:(\theta;R,t)\in \Theta\times Q\rightarrow \R$ (and designed such that $\hat{f}:(\theta;R,0)=f_0(R)$) by minimizing
        \begin{eqnarray*}
          \mathcal{L}^{(\textrm{PINN})}(\theta) = \big\|\partial_t\hat{f}-\nabla_R^2\hat{f} + \nabla_R(V_0\hat{f})-dt S_0\hat{f} \big\|^2_{L^2(Q)} \, .
        \end{eqnarray*}
In principle, the computation is performed once for all over $Q$, hence avoiding the use of N-VMC at intermediate times $t_i$. In practice, it may still however be necessary to decompose the time-domain $[0,T]$ in several subdomains ($[T_i,T_{i+1}]$ of length larger than $dt$) used in DMC to avoid the computation of a null network. 

\section{Conclusion}\label{sec:conclusion}
        NQS, {\tt Ferminet} and PINN, Deep-Ritz have become very popular libraries and algorithms which were developed independently to each other in different fields of research. In this manuscript we have established non-trivial bridges between these different methodologies.  We think that these connections may be beneficial for future developments in the different involved scientific communities. Naturally, PINN algorithms for solving \eqref{EqF} are more computationally complex than DMC-like algorithms, as they approximate PDE solutions, unlike DMC algorithms which only evaluate ``trajectories''. In future works, we plan to develop efficient PINN-like algorithms for eigenvalue problems in the spirit of DMC algorithms. Data-driven PINN algorithms will be explored for high-dimensional quantum chemistry evolution problems (for instance in the spirit of \cite{carleo3}), such as laser-molecule interactions.

 \bibliographystyle{plain}
 \bibliography{refs}
\end{document}